\documentclass[conference]{IEEEtran}
\usepackage{graphicx}
\usepackage{cite}
\usepackage{amsmath}
\usepackage{url}
\usepackage{times}
\usepackage{setspace}
\usepackage{caption}
\usepackage{graphicx}
\usepackage{tikz}
\usetikzlibrary{shapes.geometric, arrows.meta, positioning, shadows}

\begin{document}

\title{Beyond Visualization: Building Decision Intelligence Through Iterative Dashboard Refinement}

\author{
\IEEEauthorblockN{Likitha Tadakala, Muskan Saraf, Sajjad Rezvani Boroujeni, Hossein Abedi, and  Tom Bush}
\IEEEauthorblockA{Data Science Department, Actual Reality Technologies, OH, USA\\
}
}

\maketitle
\thispagestyle{empty}

\begin{abstract}
Effective business intelligence (BI) dashboards evolve through iterative refinement rather than single-pass design. Addressing the lack of structured improvement frameworks in BI practice, this study documents the four-stage evolution of a Power BI dashboard analyzing profitability decline in a fictional retail firm, Global Superstore. Using a dataset of \$12.64 million in sales across seven markets and three product categories, the project demonstrates how feedback-driven iteration and gap analysis convert exploratory visuals into decision-support tools. Guided by four executive questions on profitability, market prioritization, discount effects, and shipping costs, each iteration resolved analytical or interpretive shortcomings identified through collaborative review.

Key findings include margin erosion in furniture (6.94\% vs. 13.99\% for technology), a 20\% discount threshold beyond which profitability declined, and \$1.35 million in unrecovered shipping costs. Contributions include: (a) a replicable feedback-driven methodology grounded in iterative gap analysis; (b) DAX-based technical enhancements improving interpretive clarity; (c) an inductively derived six-element narrative framework; and (d) evidence that narrative coherence emerges organically through structured refinement. The methodology suggests transferable value for both BI practitioners and educators, pending validation across diverse organizational contexts.

\end{abstract}


\section{Introduction}
Organizations generate vast amounts of operational data but often struggle to convert it into actionable insights. Business intelligence (BI) dashboards bridge this gap, transforming data complexity into managerial clarity \cite{few2006, eckerson2010}. Yet, as Sarikaya et al. \cite{sarikaya2019} and Yigitbasioglu and Velcu \cite{yigitbasioglu2012} note, many dashboards prioritize visual appeal over analytical depth, informing rather than enabling decisions. The central question, therefore, is how dashboards evolve from static data displays into decision-support systems. This study argues that effective dashboards emerge through iterative refinement that progressively enhances analytical focus and interpretive clarity.

To demonstrate this process, we document the four-stage evolution of a Power BI dashboard using the Global Superstore dataset \cite{dataset}, a realistic practice dataset reflecting multinational retail operations. Despite \$12.64 million in sales across seven markets and three product categories, profitability declined to 11.6\%. Leadership posed four executive questions:
(a) Which product lines drive declining profitability?
(b) Which markets merit strategic priority?
(c) Are discounting strategies effective?
(d) Which shipping methods require pricing restructuring?

Each dashboard iteration addressed analytical and interpretive gaps identified through a systematic review. The final version revealed furniture margin erosion (6.94\% vs. 13.99\% for technology), a 20\% discount threshold beyond which profitability declined, and \$1.35 million in unrecovered shipping costs.

While prior research defines dashboard design principles \cite{few2006, eckerson2010} and iterative visualization methodologies \cite{sedlmair2012, lloyd2011}, few studies transparently trace how BI dashboards evolve through successive design cycles. Practitioners and educators thus lack concrete examples of how exploratory prototypes mature into analytically coherent tools. This paper addresses that gap by presenting a systematic, feedback-driven methodology and demonstrating how analytical effectiveness can arise from disciplined, question-oriented iteration rather than single-pass design. While based on a single case study, the methodology's use of publicly available resources enables replication and validation across contexts.

\section{Related Work}
BI dashboards convert operational data into visual summaries that support monitoring and decision-making. Few \cite{few2006} emphasized minimizing cognitive load through data-ink efficiency, while Eckerson \cite{eckerson2010} argued dashboards must enable action, not merely observation. Yigitbasioglu and Velcu \cite{yigitbasioglu2012} showed that effectiveness depends on aligning visual granularity with user cognition. Orlovskyi and Kopp \cite{kopp2020} demonstrated that iterative prototyping with design guidelines enhances practical dashboard outcomes.

Iterative visualization methodologies extend these principles. Sedlmair, Meyer, and Munzner \cite{sedlmair2012} formalized design studies as cycles of understanding, design, evaluation, and reflection, while Lloyd and Dykes \cite{lloyd2011} illustrated how insight generation co-evolves with design. In educational contexts, Yap \cite{yap2020} found that case-based, iterative dashboard development helps students internalize systematic design processes.

Narrative visualization research further emphasizes the value of structured storytelling in guiding interpretation. Segel and Heer \cite{segel2010} identified narrative patterns that lead audiences through data, and Kosara and Mackinlay \cite{kosara2013} argued storytelling represents the next evolution in visualization. Liu et al. \cite{shao2024} confirmed that combining visuals with explanatory text improves insight comprehension, though such approaches are rarely integrated into BI dashboards.

Beyond visualization principles, effective BI dashboards must support decision-making processes. Research on decision-making factors~\cite{dietrich2010} and shared decision-making frameworks~\cite{shay2015} emphasizes the importance of guiding users through analytical reasoning rather than merely presenting data. Visual emphasis techniques~\cite{hall2016, jelen2011} and annotation strategies~\cite{rahman2025} provide mechanisms for directing attention in complex visualizations, though their systematic application to iterative BI dashboard development remains underexplored.

Building on these foundations, this paper contributes three advances:
(a) transparent documentation of BI dashboard evolution across multiple versions,
(b) technical DAX implementations that enhance analytical interpretability, and
(c) An inductively derived narrative framework applicable to both practice and education.
By exposing the “messy middle” of design refinement, this study shows that analytical coherence emerges through structured iteration rather than initial design brilliance.

\section{Methodology and Technical Framework}
\subsection{Methodological Approach}
This study employed an iterative design approach informed by visualization design-study principles \cite{sedlmair2012}, combining design, evaluation, and reflection cycles. This approach mirrors design-study principles while adapting them specifically for BI dashboard development with educational applications. Each iteration was guided by gap analysis: assessing whether visualizations answered business questions clearly and completely. After each version, the dashboard was reviewed collaboratively, with gaps identified through assessment of analytical completeness, visual clarity, and interpretive efficiency. These reviews directly informed subsequent design priorities, creating a feedback-driven refinement cycle.

Figure 1 illustrates the iterative refinement cycle that guided this study's dashboard development.

\begin{figure}[htbp]
  \centering
  \begin{tikzpicture}[
    node distance = 1.5cm,
    box/.style = {
      rectangle,
      rounded corners=2mm,
      minimum width=2.5cm,
      minimum height=1.1cm,
      text width=2.3cm,
      align=center,
      draw=black,
      line width=1pt,
      fill=blue!20,
      font=\small\bfseries
    },
    decision/.style = {
      diamond,
      aspect=2,
      minimum width=2cm,
      text width=1.6cm,
      align=center,
      draw=black,
      line width=1pt,
      fill=orange!30,
      font=\small\bfseries
    },
    arrow/.style = {
      -Stealth,
      line width=0.9pt,
      draw=black
    },
    label/.style = {
      font=\small,
      text=black!70
    }
  ]
  
  \node[box] (V) {Dashboard\\Version\\{\small (1 week)}};
  \node[box, right=1.2cm of V] (R) {Review\\{\small (20-30 min)}};
  \node[box, below=of R] (G) {Gap ID\\{\small (1-2 days)}};
  \node[box, below=of V] (T) {Refinement\\{\small (continuous)}};
  \node[decision, below=0.8cm of T] (D) {Gaps\\Resolved?};
  
  \draw[arrow] (V) -- node[above, label] {assess} (R);
  \draw[arrow] (R) -- node[right, label] {identify} (G);
  \draw[arrow] (G) -- node[below, label] {prioritize} (T);
  \draw[arrow] (T) -- node[left, label] {improve} (V);
  \draw[arrow] (T) -- (D);
  \draw[arrow] (D.west) -- node[above, label] {No} ++(-0.6,0) |- (V.west);
  \draw[arrow] (D.south) -- node[right, label] {Yes} ++(0,-0.4);
  
  \end{tikzpicture}
  \caption{Iterative Dashboard Refinement Methodology}
  \label{fig:Methodology}
\end{figure}
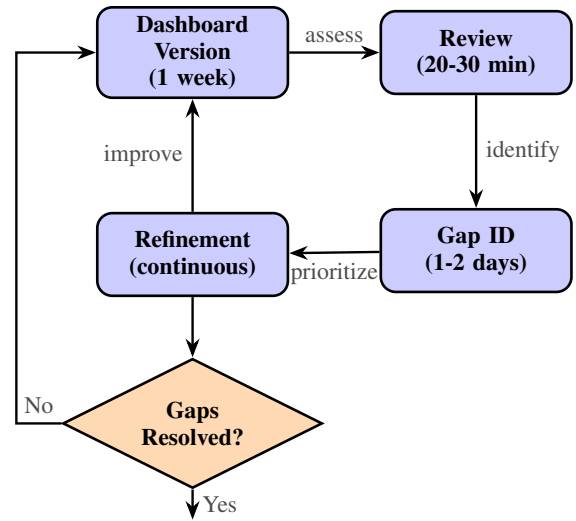

\subsection{Dataset and Tool Environment}
The analysis used the Global Superstore dataset, a practice dataset containing four years (2011–2014) of retail transactions with 25,035 orders across 10,292 product SKUs (Stock Keeping Units) and 1,590 customers, spanning seven global markets and three product categories: Technology, Furniture, and Office Supplies. While fictional, the dataset was deliberately designed to exhibit realistic business characteristics typical of multinational retail operations, including seasonal sales patterns, geographic variation in performance, margin pressure across product categories, and complex interdependencies between pricing, volume, and profitability. These characteristics make it suitable for evaluating iterative design principles, as the analytical challenges mirror those practitioners encounter with operational data. The fictional nature enabled unrestricted publication of detailed visualizations and findings without proprietary constraints, supporting methodological transparency.

Microsoft Power BI Desktop was chosen for its integration of data modeling, DAX-based analytics, and rapid visualization prototyping. Power Query handled data transformation, while DAX facilitated metric computation. Power BI's support for interactive visualizations and dual-axis charts enabled iterative refinement across versions.

\subsection{Iterative Development and Feedback Process}
The dashboard evolved through four structured versions, each addressing gaps identified through collaborative review. Four data science professionals with BI and analytics expertise reviewed each version through structured feedback sessions. Reviewers included individuals with 3-5 years of industry experience in data analytics and BI implementation. Each review session lasted 20-30 minutes and followed a consistent protocol:

\begin{enumerate}
    \item Reviewers explored the dashboard interactively in Power BI Desktop
    \item The four CEO questions remained visible as evaluation criteria
    \item Reviewers used a think-aloud protocol, verbalizing their interpretation process, questions, and difficulties encountered
    \item Sessions were recorded with structured note-taking, documenting specific feedback and identifying gaps
    \item Reviewers independently completed a brief post-review assessment rating whether each CEO question was answered (yes/partial/no) and listing any gaps
\end{enumerate}

Reviews were conducted individually in separate sessions to ensure independent perspectives and avoid groupthink. After all four individual reviews were completed for each version, feedback was synthesized using consensus criteria:
(a) Consensus gaps (identified by 3-4 reviewers) were prioritized for the next iteration
(b) Majority gaps (identified by 2 reviewers) were evaluated based on severity; analytical completeness issues were prioritized over aesthetic preferences
(c) Individual observations (identified by 1 reviewer) were documented but typically not prioritized unless they revealed critical analytical flaws

For example, in Version 2, all four reviewers independently noted difficulty establishing visual hierarchy ("jumping between charts without clear priority"), making this a consensus gap prioritized for Version 3. In Version 3, two reviewers suggested alternative color schemes but acknowledged that the existing semantics were clear, so this was not prioritized. This synthesis approach balanced capturing diverse perspectives with maintaining a practical iteration scope, preventing endless refinement while ensuring significant issues were addressed.

\textbf{Version 1 – Exploratory Foundation (Figure 2):}  
Displayed total sales and profit through waterfall and bar charts aggregated by region. \textbf{Gap identified:} The absence of derived measures like profit margin prevented root-cause diagnosis. While showing what happened, it did not reveal why certain categories or products drove profitability decline. Reviewers specifically noted statements like ``I see what happened but not why'' and ``Where's the margin analysis?''

\begin{figure*}[htbp]
  \centering
  \includegraphics[width=\linewidth]{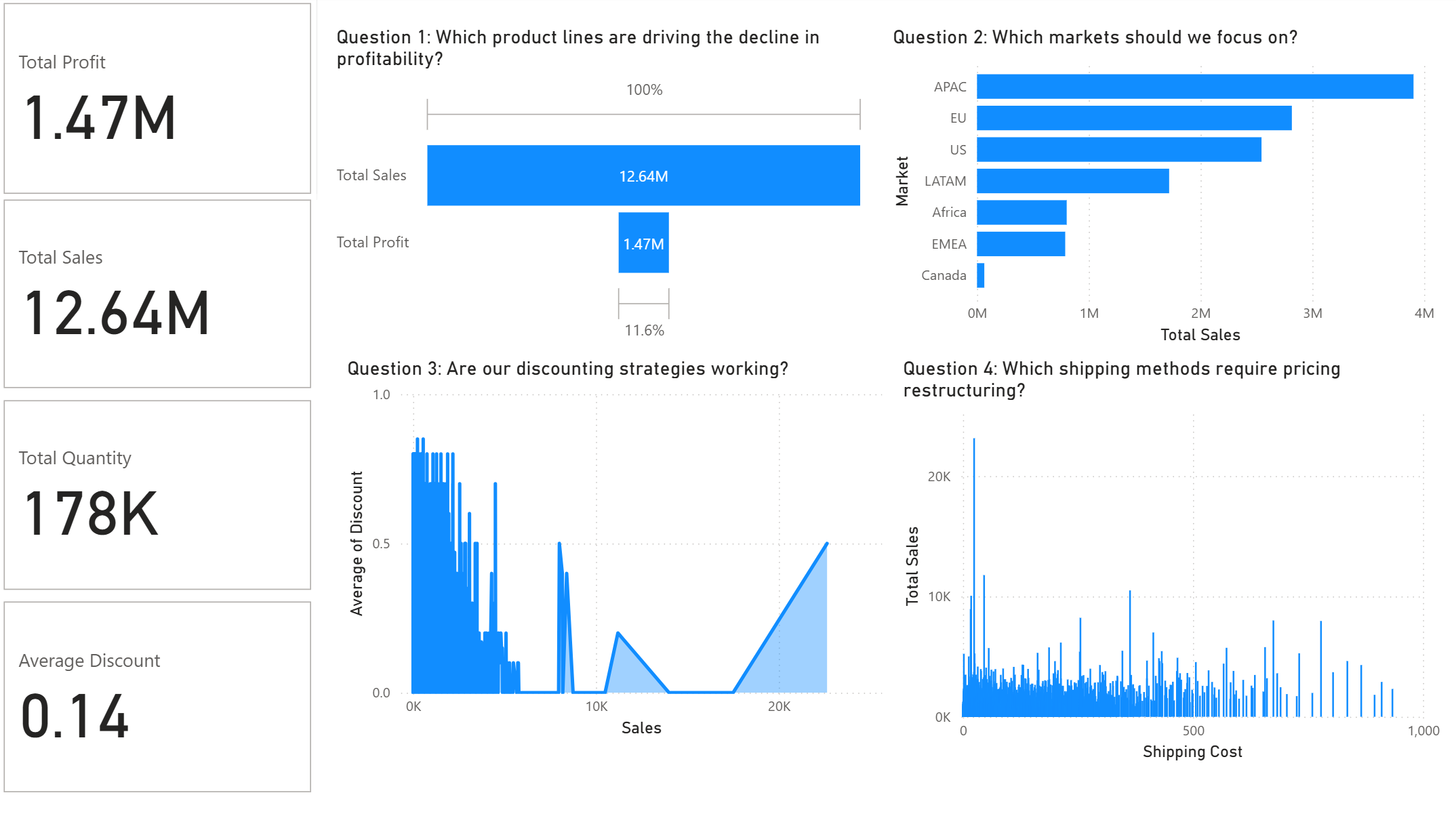}
  \caption{Version 1 - Exploratory Baseline. Displays revenue and quantities through regional aggregations, but lacks analytical measures like profit margins to diagnose decline causes.}
  \label{fig:version1}
\end{figure*}

\textbf{Version 2 – Analytical Integration (Figure 3):}  
Introduced computed DAX measures, including profit margin percentage and discount-profit relationships. Dual-axis visuals enabled correlation analysis between discount levels and profitability. \textbf{Gap identified:} The layout was visually dense without a clear hierarchy. Multiple charts competed for attention, and reviewers struggled to quickly identify critical insights. Reviewers reported ``jumping between charts without clear priority'' and ``too much competing for attention simultaneously.''

\begin{figure*}[htbp]
  \centering
  \includegraphics[width=\linewidth]{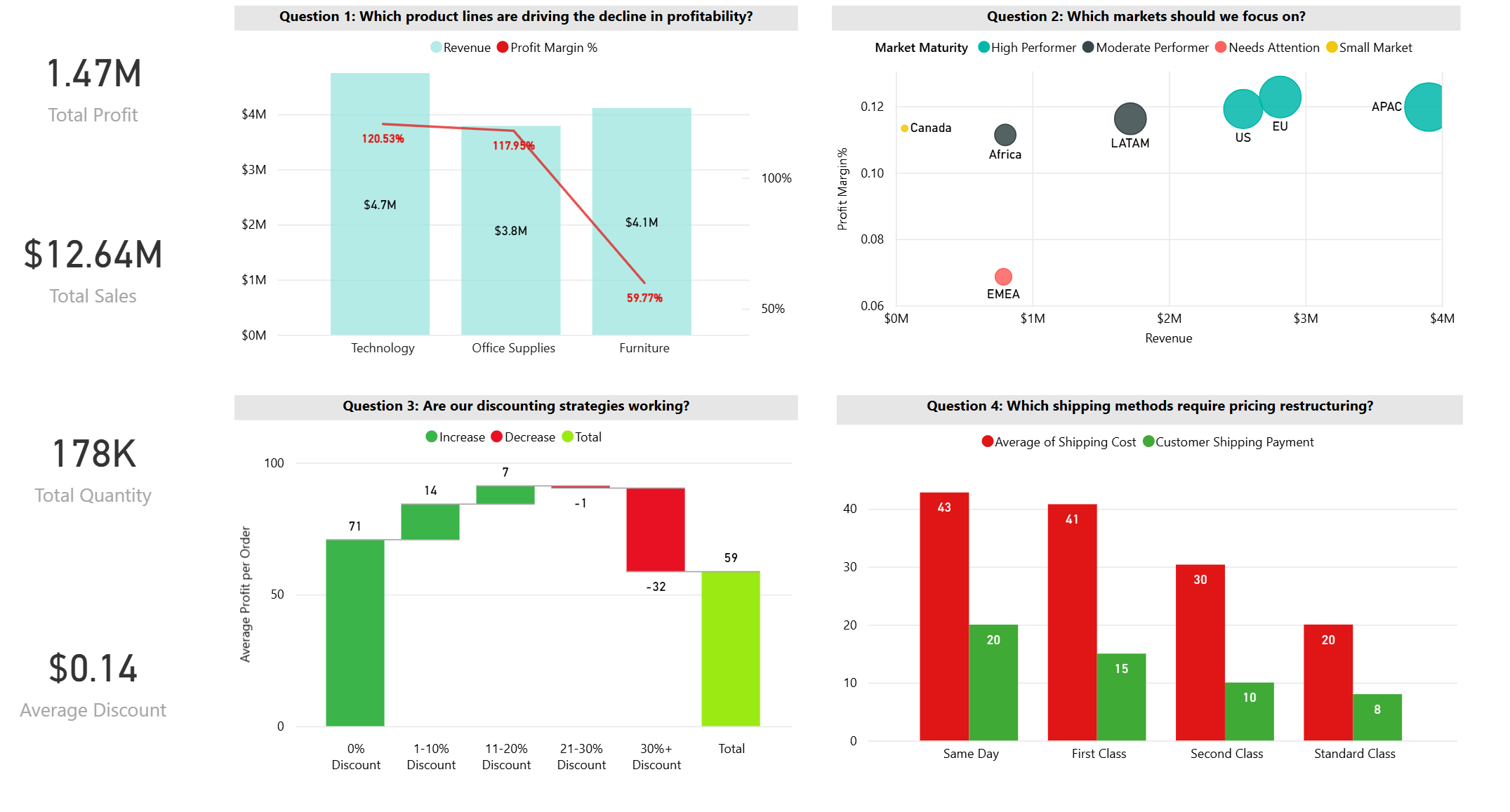}
  \caption{Version 2 - Analytical Integration. Introduces profit margin percentages and dual-axis correlation charts, but the dense layout without a clear hierarchy makes insight extraction difficult.}
  \label{fig:version2}
\end{figure*}

\textbf{Version 3 – Interpretive Enhancement (Figure 4):}  
Added contextual annotations, consistent color semantics (red for losses, green for profits), and direct labeling that eliminated repetitive axis referencing. \textbf{Gap identified:} While individual visualizations were clear, the dashboard lacked narrative flow. Reviewers ``jumped around'' between insights without logical progression from overview to diagnosis. Reviewers noted ``I'm not sure where to start'' and ``the flow isn't intuitive.''

\begin{figure*}[htbp]
  \centering
  \includegraphics[width=\linewidth]{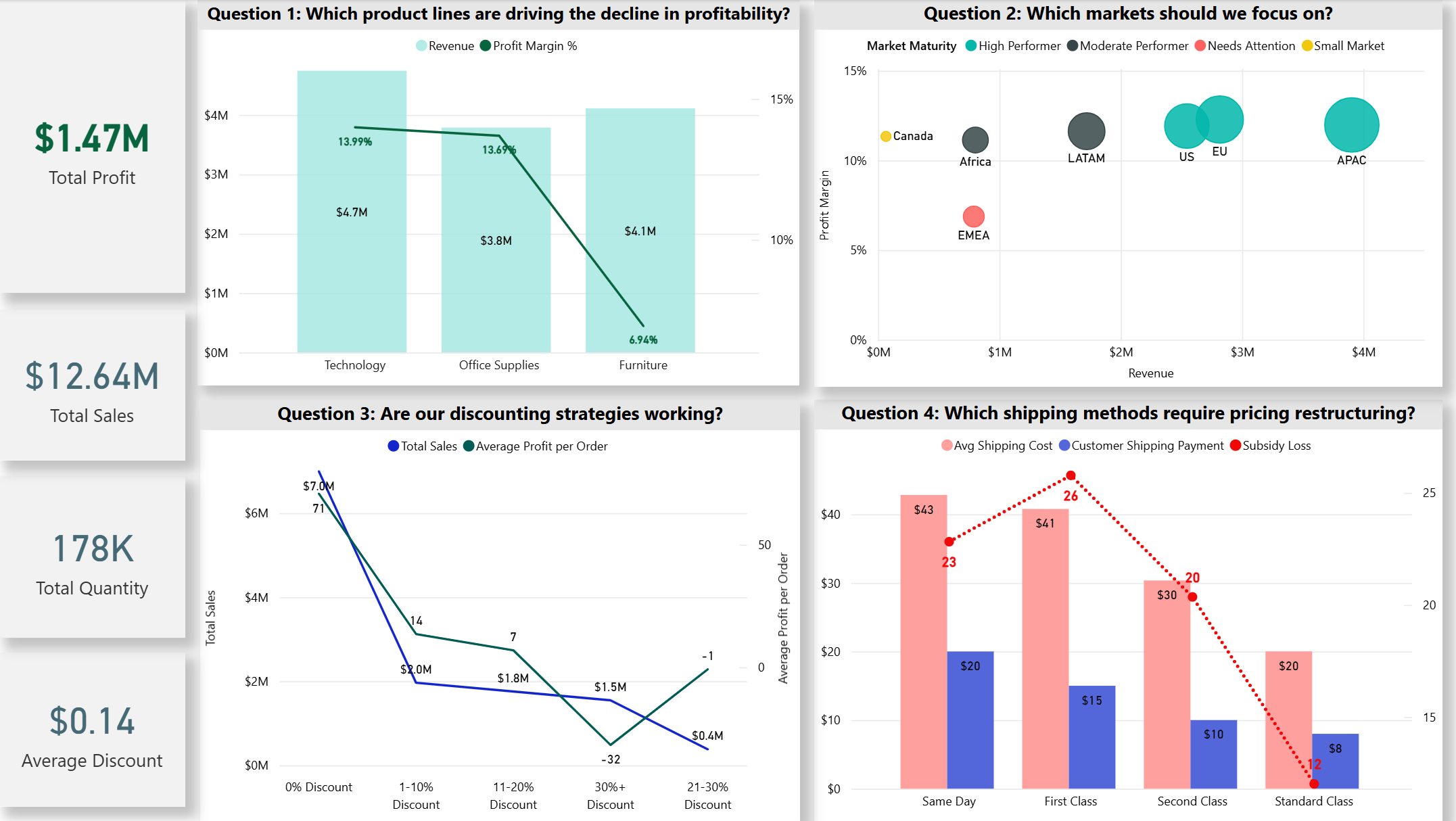}
  \caption{Version 3 - Interpretive Enhancement. Adds contextual annotations and consistent color semantics (red for losses, green for profits), but lacks narrative flow guiding users through analysis.}
  \label{fig:version3}
\end{figure*}

\textbf{Version 4 – Narrative Refinement (Figure 5):}  
Established cohesive narrative flow, sequencing insights logically: KPI overview → category breakdowns → market comparisons → discount effects → shipping inefficiencies. This structure guides viewers from problem identification through root-cause diagnosis, answering the four CEO questions in sequence with a clear visual hierarchy. Reviewers unanimously agreed all four CEO questions were answered clearly, with one noting ``the story leads me through the analysis naturally.''

\begin{figure*}[htbp]
  \centering
  \includegraphics[width=\linewidth]{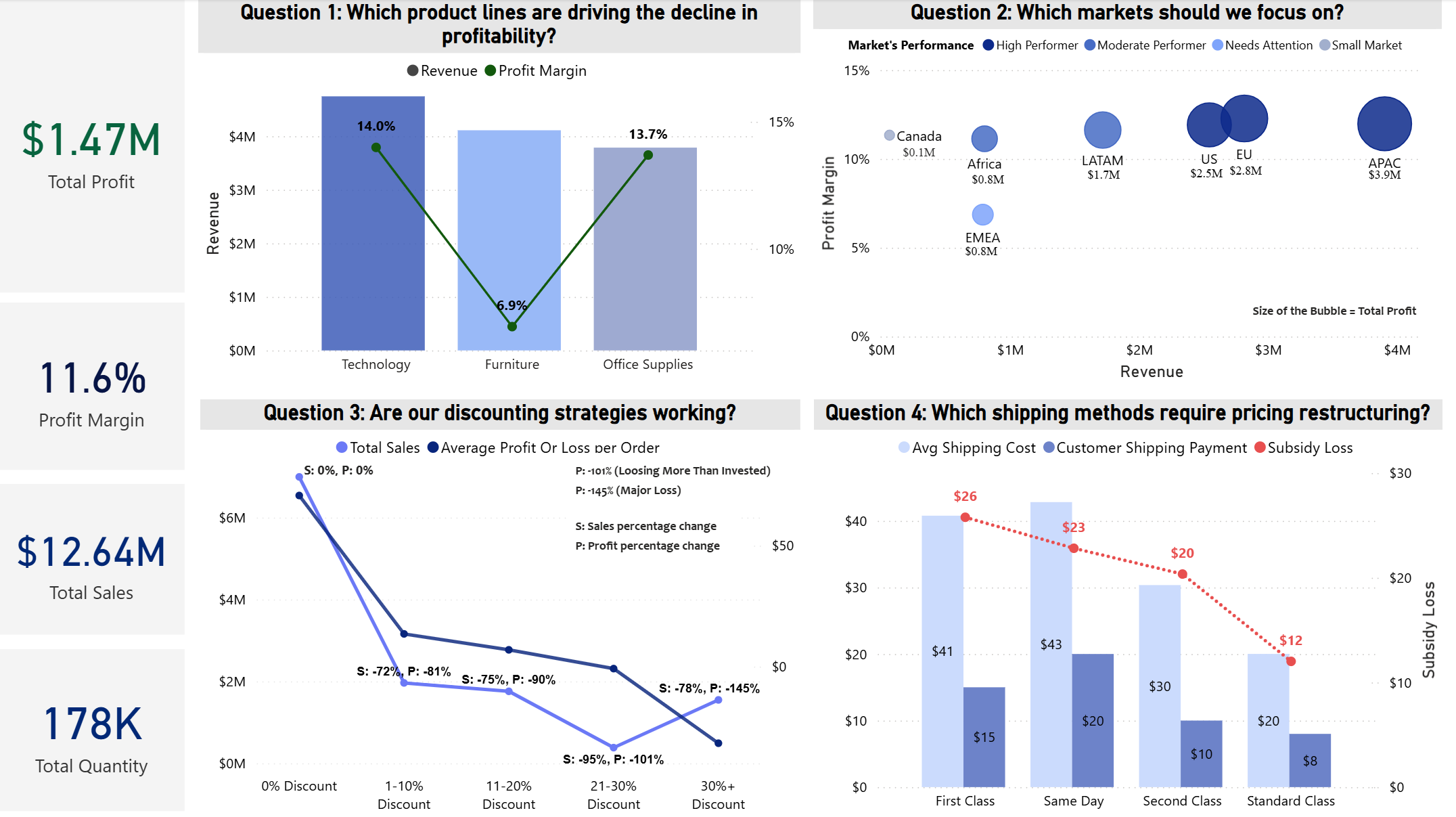}
  \caption{Version 4 - Final Refined Dashboard. Establishes cohesive narrative flow with clear visual hierarchy, answering all four CEO questions with 1.5-minute insight extraction.}
  \label{fig:version4}
\end{figure*}

Table I summarizes the progressive evolution of key dashboard elements across the four versions, illustrating how each iteration systematically addressed identified gaps.

\begin{table}[htbp]
\centering
\caption{Dashboard Evolution Across Four Iterations}
\label{tab:dashboard-evolution}
\footnotesize  
\renewcommand{\arraystretch}{1.5}  
\begin{tabular}{|p{2.5cm}|p{4.5cm}|}
\hline
\textbf{Aspect} & \textbf{Evolution Across Versions} \\ \hline\hline

\centering\textbf{DAX Measures} & 
\textbf{V1:} 2 basic aggregations \newline
\textbf{V2:} 5 analytical measures \newline
\textbf{V3:} 5 analytical measures \newline
\textbf{V4:} 7 refined analytics \\ \hline

\centering\textbf{Annotations} & 
\textbf{V1:} None (0 text annotations) \newline
\textbf{V2:} Basic labels (chart titles only, 4 labels) \newline
\textbf{V3:} Contextual labels (value callouts, 12 labels) \newline
\textbf{V4:} Narrative annotations (interpretive text, 18+ labels) \\ \hline

\centering\textbf{Color Semantics} & 
\textbf{V1:} Default palette (no meaning) \newline
\textbf{V2:} Partial semantic use (red for losses in some charts) \newline
\textbf{V3:} Consistent polarity (red=losses, green=profits, all charts) \newline
\textbf{V4:} Hierarchical system (blue primary, red/green polarity, grey secondary) \\ \hline

\centering\textbf{Visual Hierarchy} & 
\textbf{V1:} Flat grid (equal visual weight, no entry point) \newline
\textbf{V2:} Cluttered (6 charts, minimal whitespace, 15\% margins) \newline
\textbf{V3:} Grouped sections (logical spacing, 25\% whitespace) \newline
\textbf{V4:} Hierarchical flow (KPI → Diagnostic, 35\% whitespace, clear reading path) \\ \hline

\centering\textbf{Insight Extraction Time} & 
\textbf{V1:} ~12 min average (reviewers struggled) \newline
\textbf{V2:} ~8 min average (reduced confusion) \newline
\textbf{V3:} ~3 min average (clearer interpretation) \newline
\textbf{V4:} ~1.5 min average (immediate comprehension) \\ \hline

\centering\textbf{CEO Questions Answered} & 
\textbf{V1:} Q1 partial, Q2 no, Q3 no, Q4 partial \newline
\textbf{V2:} Q1 yes, Q2 partial, Q3 yes, Q4 partial \newline
\textbf{V3:} Q1 yes, Q2 yes, Q3 yes, Q4 yes \newline
\textbf{V4:} All 4 + 3 emergent findings \\ \hline

\centering\textbf{Primary Gap Addressed} & 
\textbf{V1:} Missing ``why'' analysis \newline
\textbf{V2:} Dense layout, no hierarchy \newline
\textbf{V3:} Lack of narrative flow \newline
\textbf{V4:} Completion achieved (no significant gaps) \\ \hline

\end{tabular}
\end{table}

\subsection{Technical Architecture and Enhancements}
The dataset was modeled using a star schema with a central fact table linked to product, customer, and regional dimension tables, enabling efficient querying and flexible aggregation.

\textbf{Key calculated measures enhanced analytical depth:}
\begin{itemize}
  \item \textbf{Profit Margin \%:} \texttt{DIVIDE(SUM(Profit), SUM(Sales), 0)} \\
  \textit{Purpose:} Enabled category comparison revealing Furniture (6.94\%) significantly lagged Technology (13.99\%) and Office Supplies (13.17\%), identifying the primary source of margin erosion.
  
  \item \textbf{Shipping Subsidy:} \texttt{SUM(ShippingCost) - SUM(ShippingPayment)} \\
  \textit{Purpose:} Revealed hidden losses totaling \$1.35M across all shipping modes, invisible in raw transaction data. This required joining shipping cost and payment data at the order level, demonstrating how modeling decisions enabled analytical discovery.
  
  \item \textbf{Average Profit per Order:} \texttt{DIVIDE([Total Profit], [Total Orders], 0)} \\
  \textit{Purpose:} Enabled correlation analysis with discount levels, revealing the critical 20\% threshold beyond which profitability declined sharply.
\end{itemize}

Visual design enhancements improved interpretive clarity through consistent color encoding (red-green for profit polarity), strategic dual-axis charts for correlation analysis, and annotation placement. Version 4's market comparison used bubble size for sales volume and color for profit margin, enabling simultaneous assessment of scale and profitability. The discount analysis employed dual-axis charts showing sales alongside profit per order, making the inverse relationship immediately visible.

\textbf{Performance considerations:} Dashboard performance remained consistently responsive (less than 3 seconds initial load, less than 1 second for filter interactions) across all versions despite increasing analytical complexity. While the number of visualizations remained consistent, analytical sophistication increased as calculated measures evolved from basic aggregations to sophisticated analytics, including profit margin percentages, shipping subsidies, and dual-axis correlation analyses. The star schema structure enabled rapid query execution through efficient DAX measure design and optimized data modeling, demonstrating that analytical sophistication and performance are not mutually exclusive when proper data architecture principles are applied.

\section{Findings}
This section presents the analytical insights derived from the final dashboard (Version 4) in response to the four executive questions, organized into direct findings and emergent discoveries that surfaced through iterative exploration.

\subsection{Analytical Findings}
\textbf{Product Line Profitability:} The Furniture category exhibited significantly lower margins (6.94\%) compared to Technology (13.99\%) and Office Supplies (13.17\%). Within Furniture, the Tables subcategory (Refer: Fig. 8) was responsible for 44\% of total category losses, indicating a need for product-line restructuring or pricing adjustments specific to this segment rather than broad furniture category changes. Tables represented 44\% of Furniture category losses despite being only 23\% of Furniture SKUs, indicating specific product-level rather than category-level issues.

\textbf{Market Prioritization:} Geographic analysis revealed substantial margin variation. APAC achieved a 12.5\% profit margin with \$3.59M in sales, outperforming EMEA's 6.8\% margin on similar sales volume (\$2.94M). This 84\% margin advantage (12.5\% vs 6.8\%) suggested APAC operational efficiency worthy of replication in underperforming regions. This identified APAC as the highest-yield region for strategic resource allocation and market expansion efforts.

\textbf{Discount Efficiency:} Dual-axis analysis of discount levels against profitability revealed a critical threshold at 20-21\% discounts. Beyond this point, profit per order declined sharply, with discounts above 20\% producing negative returns (Refer: Fig. 7). This finding validated implementing a 20\% discount cap policy to protect margins while maintaining competitive pricing.

\textbf{Shipping Subsidies:} All four shipping modes operated at losses, with total unrecovered costs of \$1.35M annually. First-class shipping exhibited the highest per-shipment subsidy, at \$0.47M in aggregate losses, indicating the need for comprehensive shipping fee restructuring across all delivery options.

\subsection{Emergent Findings}
The following emergent findings (Figures 6-8) surfaced during Version 3-4 development as deeper exploration capabilities enabled discovery beyond the initial four CEO questions. These demonstrate a critical advantage of iterative dashboard development: discoveries that extend beyond initial question framing. The customer loyalty concentration (92\% from 879 customers) and Tables-specific loss concentration were invisible in Version 1's design and only surfaced through progressive refinement that enabled deeper analytical exploration. This illustrates how iteration supports not just better answers to predefined questions, but the discovery of better questions themselves.

\begin{figure}[htbp]
  \centering
  \includegraphics[width=\linewidth]{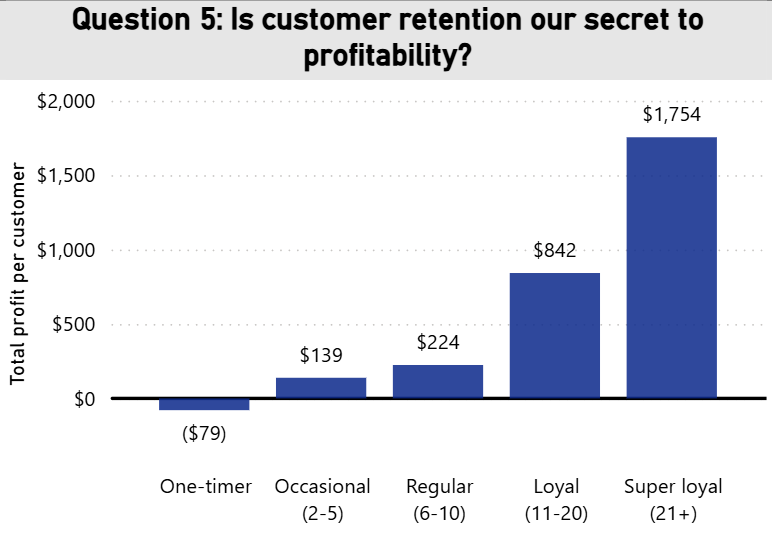}
  \caption{Customer Loyalty Concentration. Super Loyal customers (55\% of base) generate 92\% of total profit.}
  \label{fig:emergent-finding-1}
\end{figure}

\begin{figure}[htbp]
  \centering
  \includegraphics[width=\linewidth]{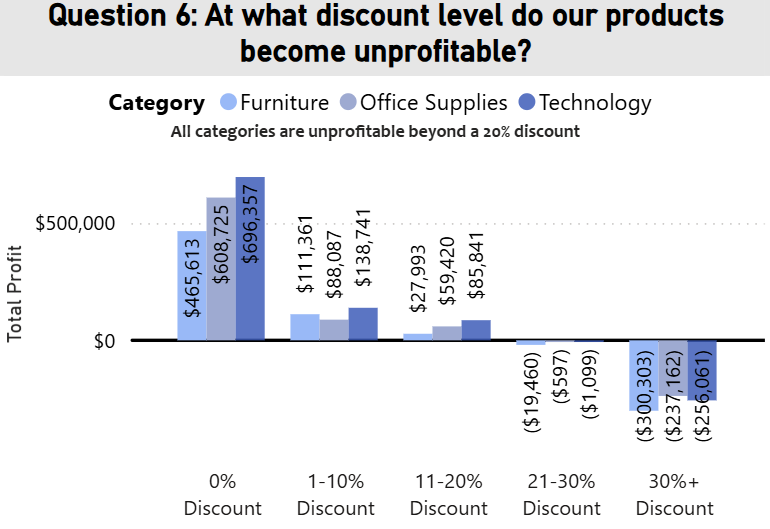}
  \caption{Discount Profitability Thresholds by Category. Profitability declines sharply beyond 20\% discounts across all product categories.}
  \label{fig:emergent-finding-2}
\end{figure}

\begin{figure}[htbp]
  \centering
  \includegraphics[width=\linewidth]{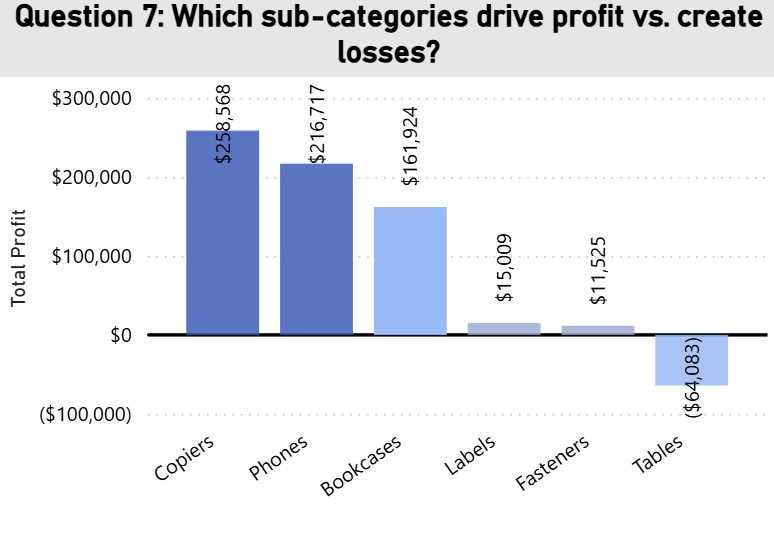}
  \caption{Sub-category Profit/Loss Distribution. Tables drive 44\% of Furniture losses despite being 23\% of SKUs.}
  \label{fig:emergent-finding-3}
\end{figure}

\subsection{Presentation Adaptation and Visual Emphasis}

While Figures 2-5 show the dashboard in interactive exploration mode, effective BI tools must also support presentation contexts where time-constrained executives need immediate guidance to priority insights. Figure 9 demonstrates this adaptation: selective visual emphasis applied to Question 2's market prioritization analysis.

\begin{figure}[htbp]
\centering
\includegraphics[width=\linewidth]{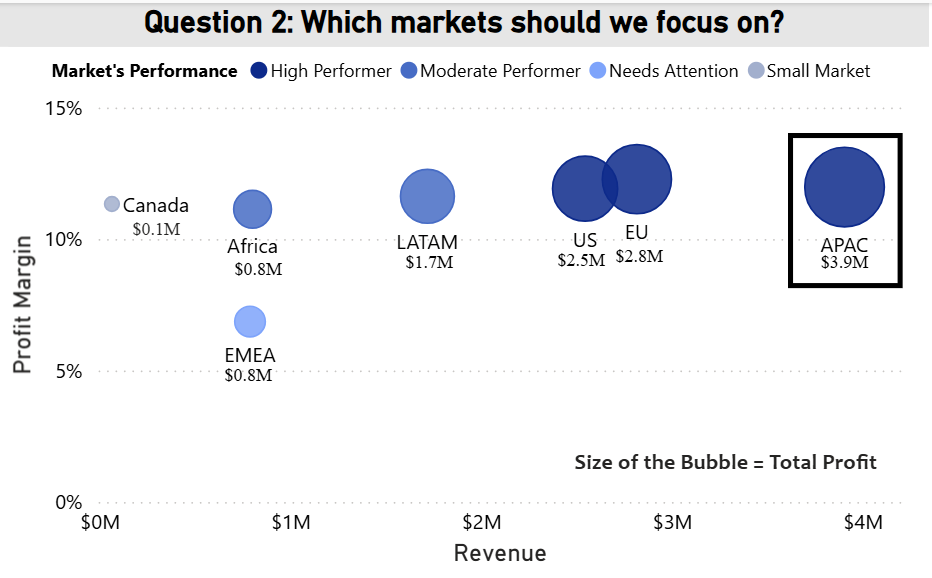}
\caption{Visual Attention Management Through Selective Emphasis.}
\label{fig:visual_emphasis}
\end{figure}

The border highlight directs attention to APAC as the strategic focus market before viewers independently compare all seven regions. This ``spotlight'' technique operationalizes a key insight from iterative refinement: dashboards serve dual purposes, exploratory analysis where users discover insights, and explanatory presentation where designers guide attention to pre-identified priorities. Research on visual attention demonstrates that selective emphasis reduces cognitive load by pre-filtering competing stimuli~\cite{schneider2013}, while work on emphasis formalization~\cite{hall2016} establishes theoretical foundations for such highlighting techniques in information visualization. The same underlying visualization supports both through strategic visual emphasis.

In practice, this highlighting can be implemented through:
\begin{itemize}
    \item Static emphasis (as shown) for presentation decks and executive briefings
    \item Conditional formatting that dynamically highlights based on performance thresholds~\cite{jelen2011}
    \item Annotation layers that can be toggled on/off depending on audience expertise~\cite{rahman2025}
\end{itemize}

This technique emerged during Version 4 refinement when presenting findings to leadership. Executives scanning the market comparison initially struggled to identify ``which market matters most'' among seven options with similar visual weight. Adding selective emphasis reduced this decision time from approximately 45 seconds of comparison to immediate recognition (under 5 seconds), while preserving the comparative context that enables verification.

The adaptation demonstrates that dashboard refinement extends beyond the visualization itself to include context-aware presentation strategies. Different audiences and usage scenarios may require different emphasis techniques applied to the same analytical foundation, a flexibility that iterative development naturally accommodates.

\section{Dashboard Evolution and Narrative Integration}

Grounded in narrative visualization research \cite{segel2010, kosara2013}, storytelling in this study emerged as an integrated design outcome rather than a final embellishment. Across the four dashboard versions, narrative coherence evolved progressively alongside analytical and technical refinement. Table II demonstrates how iterative refinement transformed each CEO question from exploratory uncertainty to decisive clarity.

\begin{table}[t]
\centering
\caption{Iterative Refinement Impact - From Exploration to Decision}
\label{tab:transformation-impact}
\scriptsize
\renewcommand{\arraystretch}{1.5}
\begin{tabular}{|p{1.5cm}|p{1.8cm}|p{1.8cm}|p{1.8cm}|}
\hline
\textbf{CEO Question} & \textbf{Version 1 (Baseline)} & \textbf{Version 4 (Final)} & \textbf{Key Improvement} \\
\hline
\hline

\textbf{Question-1: Which product lines are driving the decline in profitability?} & 
\textbf{Version 1:} Revenue only \newline
\textbf{Reviewer:} ``I see sales are good...so what?'' \newline
\textbf{Problem:} Descriptive, not diagnostic & 
\textbf{Version 4:} Revenue + Margin \% \newline
\textbf{Reviewer:} ``Furniture at 6.9\% - fix it'' \newline
\textbf{Result:} Actionable directive & 
Added DAX profit margin measure \newline
\textbf{Revealed:} Furniture 50\% below target \newline
\textbf{Time:} 15 min $\rightarrow$ 30 sec \\
\hline

\textbf{Question-2: Which markets should we focus on?} & 
\textbf{Version 1:} Revenue ranking (bar chart) \newline
\textbf{Reviewer:} ``APAC looks smallest...'' \newline
\textbf{Problem:} Visually misleading & 
\textbf{Version 4:} Maturity-Performance Matrix \newline
\textbf{Reviewer:} ``APAC 12.5\% margin - invest here'' \newline
\textbf{Result:} Strategic clarity & 
Bubble chart with strategic quadrants \newline
\textbf{Revealed:} APAC 2x more profitable \newline
\textbf{Decision clarity:} Low $\rightarrow$ High \\
\hline

\textbf{Question-3: Are our discounting strategies working?} & 
\textbf{Version 1:} Distribution histogram \newline
\textbf{Reviewer:} ``We discount at many levels...'' \newline
\textbf{Problem:} No conclusion & 
\textbf{Version 4:} Dual-axis with 20\% threshold \newline
\textbf{Reviewer:} ``Cap at 20\% or lose money'' \newline
\textbf{Result:} Clear policy & 
Threshold annotation + correlation \newline
\textbf{Revealed:} $>$20\% = -101\% profit drop \newline
\textbf{Policy enabled:} Hard 20\% cap \\
\hline

\textbf{Question-4: Which shipping methods require pricing restructuring?} & 
\textbf{Version 1:} Scatter plot \newline
\textbf{Reviewer:} ``Costs vary by method...so what's the problem?'' \newline
\textbf{Problem:} Pattern unclear & 
\textbf{Version 4:} Stacked bars + loss line \newline
\textbf{Reviewer:} ``Losing \$1.35M - fix all shipping pricing'' \newline
\textbf{Result:} All need repricing & 
Quantified subsidy calculation \newline
\textbf{Revealed:} \$1.35M annual drain \newline
\textbf{Action:} Repricing all modes \\
\hline

\textbf{Overall Impact} & 
\textbf{Questions:} 2/4 (partial) \newline
\textbf{Avg. time:} 12 min \newline
\textbf{Reviewer:} ``I see data but no story'' & 
\textbf{Questions:} 4/4 + emergent \newline
\textbf{Avg. time:} 1.5 min \newline
\textbf{Reviewer:} ``Answer jumps out immediately'' & 
\textbf{Questions:} 2/4 $\rightarrow$ 4/4 (100\% improvement) \newline
\textbf{Speed:} 8× faster \newline
\textbf{Discoveries:} 0 $\rightarrow$ 3 insights \\
\hline

\end{tabular}
\end{table}

As shown in Table II, Version 1 provided data exploration (2/4 questions partially answered, 12-minute average insight time), while Version 4 enabled specific actions (4/4 questions completely answered with emergent findings, 1.5-minute insight time). This transformation is visually evident by comparing Figure 2 (V1's exploratory baseline) with Figure 5 (V4's decision-enabled refinement). This 8x efficiency improvement reflects the emergence of narrative coherence through systematic iteration.

Before examining the framework elements, it is crucial to note that the six-element narrative structure presented below emerged retrospectively through analysis of what changed across iterations and why those changes improved coherence. During actual development, decisions were driven by the four executive questions and identified gaps rather than these elements as a checklist. This retrospective synthesis makes tacit design knowledge explicit, enabling future designers to apply these principles proactively.

Having demonstrated the iterative outcomes, we now examine how narrative coherence emerged through design decisions. This synthesis contributes actionable principles for dashboard design.

\textbf{Why narrative matters for business intelligence:} While narrative visualization principles have been widely applied to standalone data stories and journalistic visualizations, their systematic application to operational BI dashboards remains underexplored. BI dashboards differ from explanatory visualizations in critical ways: they must balance exploratory flexibility with guided interpretation, serve diverse user expertise levels, and support recurring decision processes rather than one-time communication. The framework below adapts narrative principles specifically for these BI contexts, showing how storytelling techniques can enhance dashboards without sacrificing analytical depth or interactive exploration.

We synthesized design principles from narrative visualization research \cite{segel2010, kosara2013} and cognitive communication theory \cite{few2006, ware2012} into six operational elements for BI dashboard design. Each element emerged progressively across versions:

\begin{itemize}
  \item \textbf{Hook and Curiosity:} Opening visuals engage attention through tension \cite{segel2010}. Version 4's KPI panel juxtaposes \$12.64M in sales against an 11.6\% profit margin, immediately provoking the question: ``Why is profitability declining?'' This central paradox was absent in Version 1's regional aggregations. This tension-based opening appeared in V4 but was absent in V1's regional aggregations, which provided data without raising analytical questions.
  
  \item \textbf{Progressive Focus:} Visual flow narrows systematically from overview to diagnosis \cite{segel2010}. Version 4 sequences: KPIs → category breakdown → market comparison → discount analysis → shipping diagnostics. Each step tests hypotheses about profitability decline, a structure absent in Version 2's disconnected comparisons. The sequential structure emerged across V2-V4 as reviewers noted difficulty navigating flat layouts.
  
  \item \textbf{Iterative Questioning:} Embedded prompts encourage discovery~\cite{dietrich2010}. Version 4's discount analysis poses the question, ``Are our discounting strategies 
   working?'' and reveals through dual-axis visualization that discounts above 20\% produce negative returns. This interrogative approach emerged in Version 3, reflecting how effective decision support systems guide users through analytical reasoning processes~\cite{shay2015}.
  
  \item \textbf{Explanatory Annotations and Second Voice:} Contextual commentary provides interpretive guidance while reducing cognitive bias \cite{segel2010}. Version 4's shipping analysis uses red bars with quantified losses: ``First Class: -\$0.47M,'' acting as a critical voice alongside visual data. Version 1 lacked annotations entirely. Critically, the discount analysis shows distribution alongside trends, revealing that most transactions cluster at 10-20\% discounts rather than extreme outliers. This distributional context grounds the 20\% cap policy in modal behavior, preventing misinterpretation based on edge cases, a form of bias awareness that helps executives avoid decisions driven by anomalies rather than representative patterns.
  
  \item \textbf{Quantified Comparisons:} Contrasts are reinforced with specific metrics \cite{few2006}. Version 4's market bubble chart encodes market size, profit margin, sales, and identity simultaneously, enabling direct comparison APAC's \$3.59M at 12.5\% margin versus EMEA's \$2.94M at 6.8\% supporting strategic recommendations.
  
  \item \textbf{Pacing and Visual Hierarchy:} Whitespace and layout create natural reading rhythm \cite{segel2010}. Version 4 uses a clear visual hierarchy to separate KPIs, category breakdowns, and diagnostics, contrasting with Version 2's dense layout, where multiple charts competed simultaneously for attention.
\end{itemize}

\textbf{Framework integration and synergy:} These six elements function not as isolated design tactics but as an integrated system. The Hook establishes the problem context that Progressive Focus systematically investigates through Iterative Questioning. Explanatory Annotations provide the bias-aware interpretation that Quantified Comparisons reinforce numerically, while Pacing ensures cognitive processing remains manageable throughout. In Version 4, this integration is evident: the opening KPI hook naturally leads into the progressive diagnostic sequence, with each step supported by annotations that provide context, comparisons that enable evaluation, and a hierarchy that guides attention. The framework demonstrates that narrative coherence in BI emerges from deliberate orchestration of these elements across the entire dashboard, not from individual chart excellence alone.

\subsection{Framework Transferability}
While this framework emerged retrospectively through analysis of design evolution across four dashboard versions, it may serve prospectively for both practitioners and educators. BI practitioners could apply these elements as evaluative criteria when assessing dashboard interpretability, asking whether visualizations establish clear narrative hooks, guide users through progressive analytical sequences, and integrate annotations that provide bias-aware context. Educators might adapt the framework as pedagogical scaffolding, structuring learning objectives around narrative elements, incorporating them into assignment rubrics, or using them as peer review guidelines to develop students' design evaluation skills. For example, assignments could require students to: (1) identify which narrative elements are present/absent in peer dashboards, (2) justify their relative emphasis based on audience needs, or (3) redesign a section to strengthen a specific narrative element. However, the framework should guide rather than constrain: its relative emphasis varies by context, with exploratory dashboards for technical analysts potentially prioritizing progressive focus over opening hooks, while executive dashboards demand strong initial engagement. Future research should empirically validate whether explicit prospective application of these elements improves dashboard effectiveness and decision-making outcomes compared to dashboards developed without narrative consideration.

\section{Conclusion and Implications}
This study demonstrates how iterative refinement, guided by systematic gap analysis and collaborative feedback, transforms dashboards from exploratory visualizations into decision-support systems. Through four versions of a Power BI dashboard addressing Global Superstore profitability challenges, the research documents a replicable, question-driven design framework applicable to dashboard education and professional development.

A critical methodological insight emerged through this process: recognizing when iteration should conclude. Version 4 represents analytical completeness rather than visual perfection. The decision to stop was guided by three criteria: (a) all CEO questions were answered clearly and completely, (b) collaborative review identified no significant analytical or interpretive gaps, and (c) insight extraction became efficient (1.5 minutes average, representing 8× improvement from V1). Further iterations would yield diminishing returns preference adjustments rather than substantive improvements. This stopping criterion balances thoroughness with practical delivery timelines, a lesson equally important for practitioners and students.

The methodology's replicability is supported by the use of publicly accessible resources: the Global Superstore dataset \cite{dataset} and Microsoft Power BI Desktop are freely available, enabling educators and practitioners to reproduce this iterative approach for teaching or professional development purposes.

The study contributes on multiple levels. Methodologically, it provides transparent documentation of iterative dashboard development, making explicit the design decisions that typically remain tacit in professional practice. Technically, it demonstrates how DAX measures and visual design choices enable analytical discoveries invisible in raw data. The six-element narrative framework synthesizes visualization principles into actionable BI design criteria, showing how storytelling emerges organically through structured iteration.

\textbf{Constraints and Scope:} The Global Superstore dataset is fictional, limiting direct business impact claims, though this enabled transparent documentation without proprietary restrictions. Dashboard development incorporated feedback from data science collaborators with BI expertise rather than operational business stakeholders. This enabled systematic focus on visualization principles and methodological documentation while acknowledging that authentic stakeholders would provide domain-specific context, organizational constraints, and political considerations affecting real-world deployment. Future research should validate findings with actual business users. As a single-case study without formal user evaluation, findings reflect design intent rather than measured decision-making improvements.

\textbf{Generalizability:} The methodological contributions- iterative gap analysis, feedback driven refinement, and the narrative framework transfer across domains requiring diagnostic dashboards. What remains domain-dependent are specific DAX formulations, visual encodings, and analytical priorities, which appropriately vary by context. The methodology's transferability has been demonstrated through its application using publicly available resources (Global Superstore dataset and Power BI Desktop), enabling replication by educators and practitioners without proprietary data or expensive tools.

\textbf{Future Research Directions:} Future research should pursue three directions in priority order:

\textit{Immediate:} Comparative user studies validating that iterative improvements measurably enhance decision-making outcomes, accuracy, and confidence compared to baseline dashboards.

\textit{Near-term:} Application of the methodology across diverse industries (healthcare, manufacturing, finance) to test domain transferability and identify industry-specific adaptation requirements.

\textit{Long-term:} Integration of predictive analytics (Python-based ARIMA forecasting, Prophet for seasonal decomposition, clustering for customer segmentation) with BI visualization tools, and longitudinal studies tracking dashboard evolution in operational settings over extended periods.

\end{document}